\documentclass[
    ,final            
  ]
  {aipproc}

\layoutstyle{8x11double}

\usepackage{psfrag}
\usepackage{bm}
\newcommand{\ld}{\ensuremath{{\cal{L}}}}
\psfrag {LD2} {$\bm \ld$}
\newcommand{\ttbar}{\ensuremath{t\bar{t}}}
\newcommand{\ppbar}{\ensuremath{p\bar{p}}}
\newcommand{\Dzero}{D{\O}}
\newcommand{\zprime}{\ensuremath{Z^{\prime}}}
\newcommand{\pt}{\ensuremath{p_{T}}}
\newcommand{\met}{\ensuremath{\not\!\!{E}_{T}}}

\begin{document}

\title{Measurement of the Forward-Backward Charge Asymmetry in Top-Quark Pair Production in Proton-Antiproton Collisions at \Dzero}


\author{Su-Jung Park on behalf of the \Dzero\ Collaboration}{
	address={II. Physikalisches Institut,
		Georg-August-Universit\"at G\"ottingen \\
		Friedrich-Hund-Platz 1,
		37077 G\"ottingen,
		Germany}}

\begin{abstract}
A measurement of the forward-backward charge asymmetry in top-antitop 
(\ttbar) pair production in proton-antiproton (\ppbar) collisions is 
presented. The asymmetry is measured for different jet multiplicities in 
the lepton+jets final state on 0.9~fb$^{-1}$ of data collected by the
\Dzero\ experiment at the Fermilab Tevatron Collider. The result is 
sensitive to new physics, which is demonstrated by setting an upper limit 
on \ttbar\ production via heavy neutral gauge bosons (\zprime).
\end{abstract}

\maketitle


\section{Introduction} \label{sec_intro}
At the Tevatron, top quarks are mainly produced in top-antitop-quark 
(\ttbar) pairs. Since the initial proton-antiproton (\ppbar) state is
not a charge conjugation ($C$) eigenstate, the final state is not expected
to be symmetric under $C$ either.


The charge asymmetry is therefore not expected to vanish and can be
translated to a forward-backward asymmetry.
Because of limited statistics, instead of a differential measurement
this analysis \cite{amnon} measures the integrated forward-backward asymmetry:
\begin{equation}
A_{fb} = \frac{N_f-N_b}{N_f+N_b},
\end{equation}
where $N_f$ ($N_b$) is the number of forward (backward) events, meaning events,
where the rapidity difference $\Delta y = y_t - y_{\bar{t}}$ between top quark
and antitop quark
is positive (negative).

At leading-order (LO) \ttbar\ production is charge symmetric. However, starting
at next-to-leading-order (NLO), an asymmetry arises from the inferences between
production diagrams. A total asymmetry of 4--5\% is expected \cite{asym1}. The
contributions to $A_{fb}$ differ depending on which processes interfere. They
are positive (6.4\% NLO) for $2 \to 2$ processes while they are negative 
(-(7--8)\% NLO, -(3--0)\% NNLO) for $2 \to 3$ processes, which result in at 
least one more jet in the final state \cite{asym2,asym3}. By measuring the 
asymmetry for different jet multiplicities, this analysis attempts to verify 
this behavior.



\section{Measurement of the Forward-Backward Charge Asymmetry} \label{sec_search}
The presented analysis uses approximately 900~pb$^{-1}$ of data collected by
the \Dzero\ experiment \cite{D0RUNII} at Fermilab and proceeds with the 
following steps. First a sample
enriched in \ttbar\ events is selected, then each event is fully reconstructed 
using a kinematic fitter. After that, the sample is fitted for sample 
composition and asymmetry simultaneously. To suppress model-dependent effects,
the result is not corrected for acceptance, instead a simple description of the 
accepted phase space is given. Likewise, it is not corrected for reconstruction 
effects, but a dilution is given to be used with any model.

\begin{figure}[h!]
	\includegraphics*[width=0.45\textwidth]{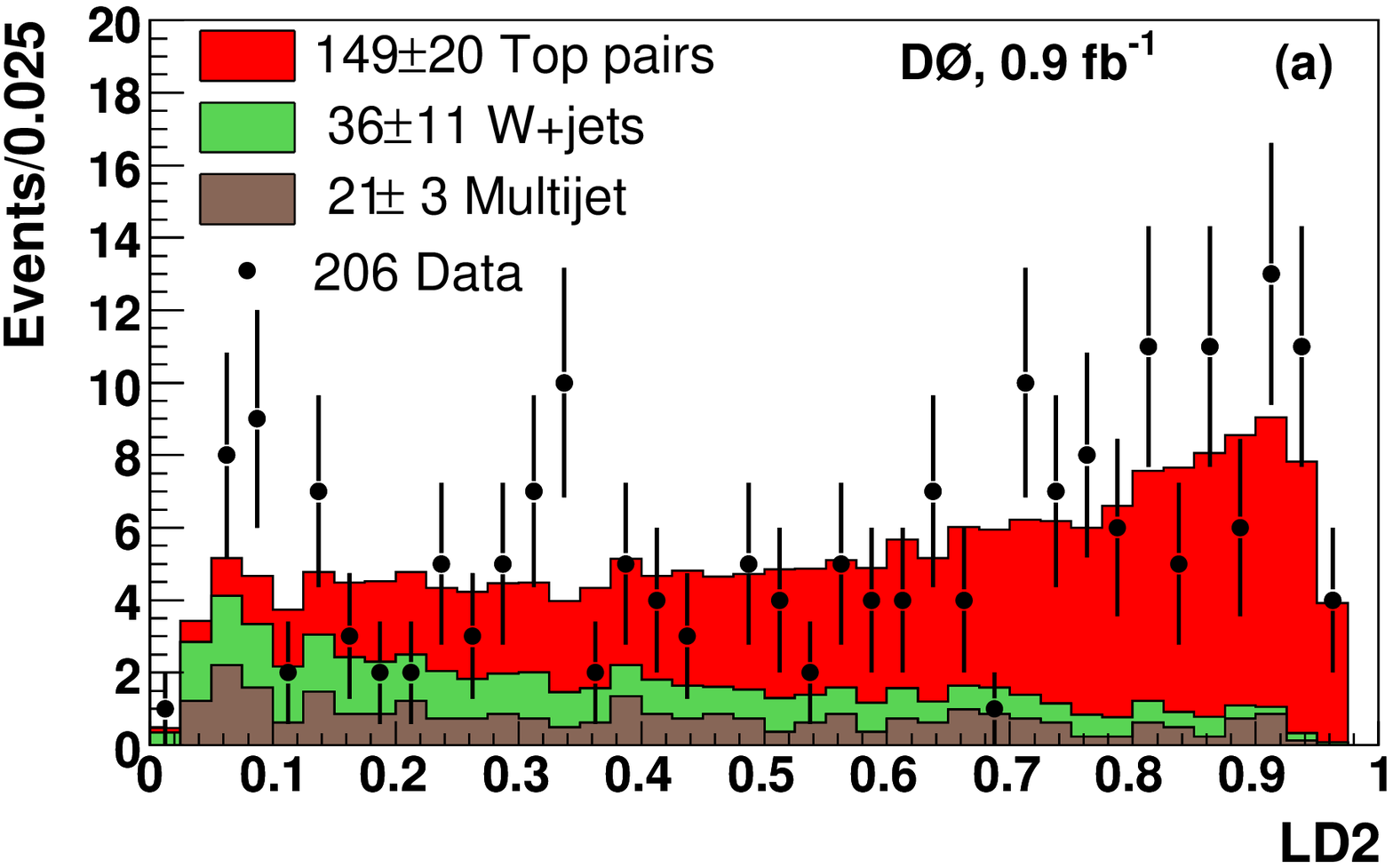}
	\includegraphics*[width=0.45\textwidth]{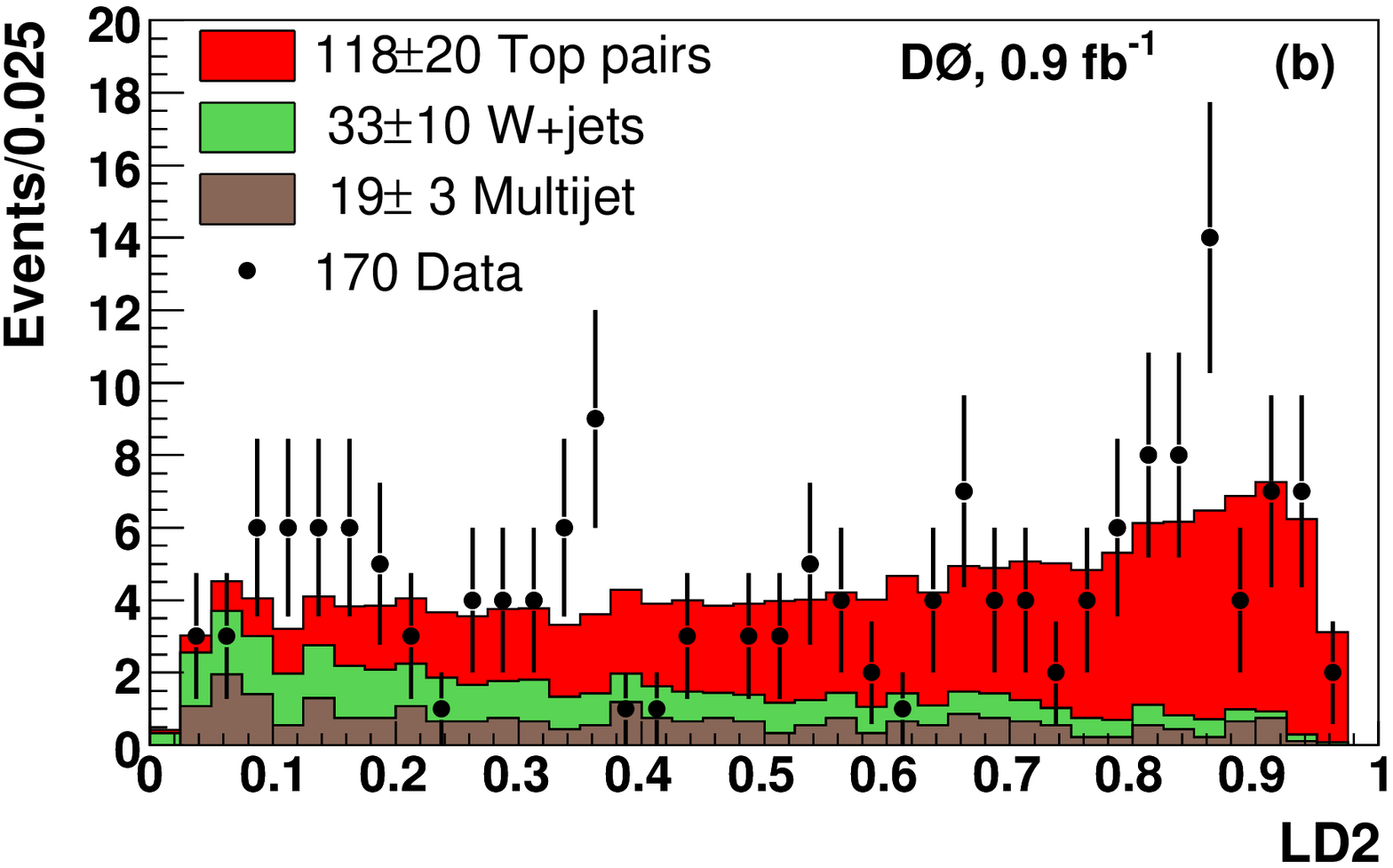}
	\caption{\label{fig_recoy} Fitted likelihood discriminant for
	events with $\ge$4 jets. Events reconstructed (a) as forward
	($\Delta y>0$) and (b) as backward ($\Delta y<0$).}
\end{figure}

\subsection{Event Selection}
Each the top and the antitop quark decays to a $W$~boson and a $b$ quark.
The subsequent decay of the $W$~boson determines the event topology. 
For this analysis, we consider the decay channel with one $W$~boson decaying 
to hadrons and the other one to leptons. The resulting final state consists 
of one high-\pt\ lepton, missing transverse energy \met\ from the neutrino, 
two jets originating from a long-lived $b$ hadron ($b$ jet), and two light 
quark jets. This will be referred to as the lepton+jets channel. 

Based on the final state, events with $\ge$4 reconstructed jets are
selected. The jets have to have transverse momentum $\pt>20$~GeV (35~GeV for 
the leading jet) and pseudorapidity $|\eta|<2.5$. At least one of the jets
has to be identified as a $b$ jet by a neural network tagging algorithm 
\cite{btag}. We also require exactly
one isolated electron with $\pt>15$~GeV and $|\eta|<1.1$ or one isolated
muon with $\pt>18$~GeV and $|\eta|<2.0$. The event has to have a primary
vertex, which fulfills certain quality requirements, and $\met>15$~GeV.

After the selection, each event is reconstructed to a \ttbar\ hypothesis
by a kinematic fitter \cite{hitfit}. It varies the four-momenta of each 
object within their resolutions and minimizes a $\chi^2$ statistic under
the constraint that the $W$~boson masses are exactly 80.4~GeV and the top
quark masses exactly 170~GeV. The $b$-tagging information is used to
reduce the jet combinatorics.

Although it is also possible to determine the aysmmetry using only the
lepton charge and direction, reconstructing the whole event doubles
the sensitivity.

\subsection{Sample Composition and Asymmetry}
The largest background processes to \ttbar\ production are $W$+jets and 
multijet production. The size and asymmetry of the multijet background is
estimated from data, using a sample with non-isolated leptons. To be able
to discriminate from the $W$+jets background, a likelihood discriminant 
$\cal{L}$ is defined, which is based on variables which are well-modeled
by Monte Carlo, provide good separation between \ttbar\ and $W$+jets, and
do not bias the rapidity difference $|\Delta y|$. The asymmetry of the 
$W$+jets background is suppressed by the kinematic fit, and the remaining
asymmetry is estimated by simulation. The other small sources of background 
are taken into account in the systematic uncertainties. The likelihood
distributions are fitted simultaneously for sample composition and sign of
$\Delta y$ using a maximum likelihood method. Figure~\ref{fig_recoy} shows
the outcome of the fit for events with $\ge$4 jets in data and Monte Carlo.

The resulting asymmetry is $A_{fb}$ = [12 $\pm$ 8(stat.) $\pm$ 1(sys.)]\% for 
events with $\ge4$ jets. Table~\ref{tab_res} shows the asymmetry as it 
was measured on data and as predicted by the generator MC@NLO \cite{mcatnlo} for 
different jet multiplicities.
\begin{table}[h!tbp]
	\caption{\label{tab_res}
	Asymmetry in percent measured on data and as predicted by MC@NLO for 
	different jet multiplicities.}
	\begin{tabular}{crr}
	\hline 
	\tablehead{1}{c}{c}{$N_{jet}$} & \tablehead{1}{c}{c}{\Dzero\ data} & \tablehead{1}{c}{c}{MC@NLO} \\
	\hline 
	$\ge4$ & 12$\pm\;$8(stat)$\pm$1(sys) & 0.8$\pm$0.2(stat)$\pm$1.0(acc) \\
	\hline
	$=4$   & 19$\pm\;$9(stat)$\pm$2(sys) & 2.3$\pm$0.2(stat)$\pm$1.0(acc) \\
	$\ge5$ & -16$^{+15}_{-17}$(stat)$\pm$3(sys) & -4.9$\pm$0.4(stat)$\pm$1.0(acc) \\
	\hline 
	\end{tabular}
\end{table}

\subsection{Acceptance and Reconstruction Effects}
\begin{figure}[h]
	\includegraphics*[width=0.45\textwidth]{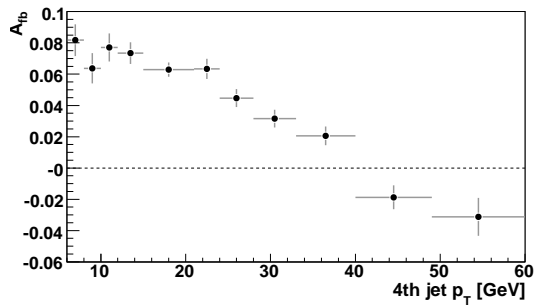}
	\caption{\label{fig_acceptance} The asymmetry as predicted by MC@NLO
as a function of the 4th highest particle \pt.}
\end{figure}

The integrated asymmetry strongly depends on the region of phase space probed.
The largest effect stems from requiring $\ge$4 jets above a certain \pt\
threshold, which is strongly correlated to the 4th highest particle jet \pt.
Figure~\ref{fig_acceptance} shows how the generated asymmetry depends on that
variable.

Since the exact structure of the asymmetry in the Standard Model is unknown,
we choose not to correct for the acceptance. Instead, the analysis is designed
so that the selection can be described by simple particle-level cuts. This
description has been shown to yield an accuracy of 2\% (absolute).

\begin{figure}[h]
	\includegraphics*[width=0.45\textwidth]{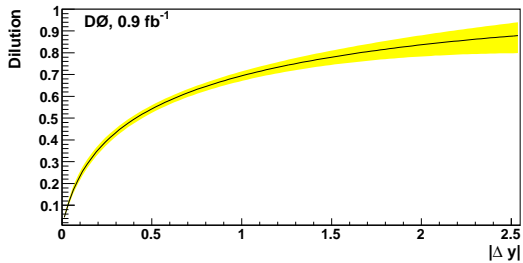}
	\caption{\label{fig_dilution} The dilution and its uncertainty as
	a function of generated $|\Delta y|$ for \ttbar\ events with $\ge$4 jets.}
\end{figure}

Misreconstruction of the sign of $\Delta y$ dilutes the observed asymmetry.
It can occur by misidentifying the lepton charge, a negligible effect, and 
by misreconstructing the event geometry. If $p$ is the probability to measure
the sign of $\Delta y$ correctly, the fraction of visible asymmetry, which is
called the dilution factor, is $D = 2p - 1$. Since corrections for the 
reconstruction effects are highly model-dependent, the parametrization of the
dilution for different jet multiplicities is provided instead. It is shown for 
events with $\ge$4 jets in Figure~\ref{fig_dilution}.

To compare a predicted asymmetry from any model with the observed one,
the generated $A_{fb}$ has to be folded with the acceptance and reconstruction
effects. This was done with the MC@NLO prediction to obtain the numbers in
Table~\ref{tab_res}.

\section{Sensitivity to New Physics}
\begin{figure}[h]
	\includegraphics*[width=0.45\textwidth]{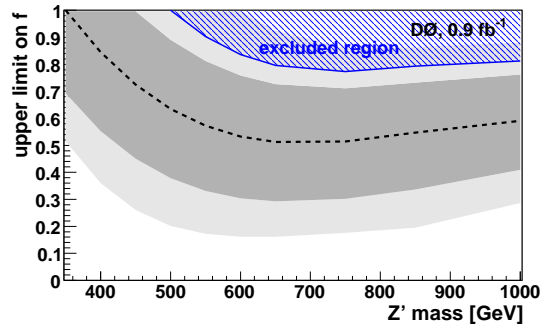}
	\caption{\label{fig_zprime} 95\% C.L.\ limits on the fraction of \ttbar\
	produced via a \zprime\ resonance as a function of the \zprime\ mass. The
	dashed curve indicates the expected limit with the shaded bands at one and
	two standard deviations. The solid curve represents the observed limit.}
\end{figure}

Resonance production of \ttbar\ can change the forward-backward charge 
asymmetry. Axigluons are predicted to give negative asymmetries \cite{axigluons}.
Production via heavy neutral gauge bosons (\zprime), however, can give large
positive contributions to the asymmetry, and this example was studied in this
work. By measuring the asymmetry one is in principle sensitive to narrow 
and wide resonances. The distribution for $A_{fb}$, as a function of the fraction 
$f$ of \ttbar\ events produced via a leptophobic \zprime\ with $Z$-like coupling
to quarks, was predicted for several \zprime\ masses using ensembles of simulated 
datasets. Employing the Feldman-Cousins method \cite{feldmancousins} limits at 
95\% confidence are derived and displayed in Figure~\ref{fig_zprime}.

\section{Summary} \label{sec_result}
A first measurement of the integrated for\-ward-back\-ward charge asymmetry in top 
quark pair production was presented. The measured asymmetries are consistent with
the MC@NLO predictions, and the expected behavior for events with $=4$ jets and
$\ge5$ jets is indicated by the data. Parametrizations of the acceptance and dilution
allow comparisons with any model. Limits on \ttbar\ production via heavy neutral
gauge bosons can be determined.


\begin{theacknowledgments}
We acknowledge support from the Bundesministerium f\"ur Bildung und Forschung (BMBF)
in Germany.
\end{theacknowledgments}


\bibliographystyle{aipproc}   

\end{document}